\documentclass[showpacs,altaffilletter,preprintnumbers,aps,twocolumn,amsmath,amssymb]{revtex4}
\usepackage{graphicx}
\usepackage{dcolumn}
\usepackage{bm}

\begin{document}
\title{The role of Joule heating in the formation of nanogaps by electromigration}

\author{M.L. Trouwborst}
\email{M.L.Trouwborst@rug.nl}

\author{S.J. van der Molen}
\altaffiliation{Present address: Nanocenter Basel, Institute of
Physics, Klingelbergstrasse 82, CH-4056, Basel, Switserland }

\author{B.J. van Wees}
\affiliation{Physics of Nanodevices, Materials Science Centre,
Rijksuniversiteit Groningen, Nijenborgh 4, 9747 AG Groningen, The
Netherlands}
\date{\today}
\begin{abstract}

We investigate the formation of nanogaps in gold wires due to
electromigration. We show that the breaking process will not start
until a local temperature of typically 400 K is reached by Joule
heating. This value is rather independent of the temperature of
the sample environment (4.2-295 K). Furthermore, we demonstrate
that the breaking dynamics can be controlled by minimizing the
total series resistance of the system. In this way, the local
temperature rise just before break down is limited and melting
effects are prevented. Hence, electrodes with gaps $ <$ 2 nm are
easily made, without the need of active feedback. For optimized
samples, we observe quantized conductance steps prior the gap
formation.

\end{abstract}
\maketitle
\section {introduction}

For single molecular electronics, electrodes are needed with a
separation of typically one nanometer. In recent years, some
creative methods have been developed to fabricate such devices.
Some with tunable electrode distance, like scanning probe methods
\cite{weiss, xu} and mechanically controllable break
junctions,\cite{smit, dulic, reichert} and others with a fixed
distance, like electromigration-induced tunnel
devices.\cite{liang, park, Hpark} The latter have the advantage
that they can easily be extended with a third (gate) electrode.
Also, compared to other procedures, electromigration-induced
nanogaps are easy to prepare, making this method one of the most
promising techniques.\

Electromigration-induced nanogaps are produced by applying large
current densities to gold wires. At high current densities $j$
(typically $10^{8}A/cm^{2}$), momentum transfer from the electrons
to the gold atoms leads to drift of the atoms, in the direction of
the electron flow. This mass flux can lead to the growth of voids
in the wire, finally leading to the formation of
gaps.\cite{mahadevan} These gaps can have widths as small as a
single nanometer, which makes them suitable for single molecular
electronics.\cite{Parkapl,bolotin} However, to obtain gaps smaller
than 2 nm, it is crucial that the process is indeed dominated by
electromigration. Joule heating, resulting in melting and surface
tension effects, can be the cause of much bigger gaps and gold
island formation.\cite{strachan, hubert,houck}

It is a challenging problem to prevent excessive heating. The
reason for this is that a high current density is only one of the
requirements for nanogap formation. An additional condition is
that the atomic mobility is high enough for substantial mass flux
to occur. Since the mobility shows activated behavior, the local
temperature plays a key role in electromigration. If the
temperature of the sample environment is low (e.g., 77 K), Joule
heating has to be substantial to induce the required local
temperature. Below, we show, using a simple model and local
temperature measurements, that gap formation typically takes place
at a local temperature $T_{start} \sim 400$ K. This value is
rather independent of the temperature of the sample environment
(4.2 K, 77 K, or 295 K). \\
Unfortunately, the local temperature does not stay constant once
the breaking process has started. While the slit is being formed,
the current density and, hence, the local Joule heating increases.
As a consequence, the local temperature diverges just before break
down. In fact, temperatures can be reached up to the melting point
of gold, leading to ill-defined junctions. This problem can be
solved by using an active feedback system, where the speed of the
electromigration process is kept constant.\cite{strachan, fuhrer,
herre, hubert, houck} In this way, the temperature is limited,
giving better control on the final gap size. In this article, we
discuss a simple model to describe the local temperature rise just
before break down. We test this model by doing local temperature
measurements during the electromigration process. We show that
excessive heating can be prevented, by limiting the series
resistance of the system. This makes a feedback system
unnecessary. By optimizing our sample design, we have increased
the number of junctions with gaps $< 2$ nm from 15 to $>$ 90
percent. Furthermore, constrictions of only one or a few atoms can
easily be achieved, leading to conductance steps of typically
$2e^{2}/h$.

\section{theory: Electromigration in Metals}

Electromigration is commonly described as a mass flux under the
influence of a high current density. However, a full description
of the electromigration problem is not trivial. To give some
insight, we refer to thermodynamics of irreversible processes.
This theory considers \textit{all} fluxes and forces
involved.\cite{groot} In our case, there are three types of fluxes
to be dealt with: the electron particle flux ($\bf{J_{e}}$), the
flux of metal atoms ($\bf{J_{m}}$), and the energy flux
($\bf{J_{u}}$). These fluxes are induced by a set of three
'forces' (or potential gradients), \textbf{X}$_j$. For the
particle forces, we can write $\bf{X}_j=-\nabla \mu^{j}_{ec}$.
Here, $\mu_{ec}=\mu + Z e \varphi$ is the electrochemical
potential, with $\varphi$, $\mu$ and $Z$, the electrostatic
potential, the chemical potential and particle charge (-1 for
electrons), respectively. The other 'force' is due to a
temperature gradient: $\bf{X}_u=\nabla(1/T)$. From thermodynamics
of irreversible processes we have following set of equations:

\begin{equation}\label{au}
    \textbf{J}_m=-{L_{m,m}}\nabla\left(\frac{\mu^{m}_{ec}}{T}\right)-
{L_{m,e}}\nabla\left(\frac{\mu^{e}_{ec}}{T}\right)-{L_{m,u}}\left(\frac{\nabla
T}{T^{2}}\right)
\end{equation}

\begin{equation}\label{el}
    \textbf{J}_e=-{L_{e,m}}\nabla\left(\frac{\mu^{m}_{ec}}{T}\right)-
{L_{e,e}}\nabla\left(\frac{\mu^{e}_{ec}}{T}\right)-{L_{e,u}}\left(\frac{\nabla
T}{T^{2}}\right)
\end{equation}

\begin{equation}\label{u}
    \textbf{J}_u=-{L_{u,m}}\nabla\left(\frac{\mu^{m}_{ec}}{T}\right)-
{L_{u,e}}\nabla\left(\frac{\mu^{e}_{ec}}{T}\right)-{L_{u,u}}\left(\frac{\nabla
T}{T^{2}}\right)
\end{equation}

where the phenomenological constants $L_{ij}$ relate all fluxes to
all forces, whilst obeying the Onsager relations, $L_{ij}=L_{ji}$.
The coupled equations above provide a general description of the
system. This description includes electromigration as well as
thermo-electric effects and thermodiffusion. The latter refers to
a mass flux due to temperature gradients (which can, in turn, be
due to Joule heating). Unfortunately, to solve the set of
equations for a three-dimensional geometry is a formidable task.
With a few assumptions, however, a simplified relation for the
mass flux can be obtained. First, for materials with a high
conductivity, we can ignore $\nabla \mu^e$. Second, we choose to
neglect thermodiffusion. In the experimental part of this article,
this assumption will be justified.  Finally, we note that in
practice all charge current is due to electron flux, so that
$-\nabla \varphi=\rho j$ (with $\rho$ the electrical resistivity).
Hence, we obtain:

\begin{equation}\label{fluxmetal}
    \textbf{J}_m=-L^{*}_{m,m}(\nabla \mu^m - Z^{*}e\rho j)
\end{equation}

where we defined $L^{*}_{i,j}=L_{i,j}/T$. Furthermore, we
introduced an effective charge $Z^{*}$:

\begin{equation}\label{zeff}
    Z^*=Z-\frac{L^{*}_{m,e}}{L^{*}_{m,m}}
\end{equation}

In words, the atoms behave as if they had a charge
$Z^{*}$.\cite{molen, thin film, groot} This effective charge is
due to momentum transfer from electrons to atoms. Generally,
$\frac{L_{me}}{L_{mm}}>>Z$, so the net force acting on the gold
atoms will be in the direction of the electron flow.

\section{Experimental setup}

Two kinds of samples are fabricated on top of a 500 nm thick
SiO$_{2}$ layer on Si. For the first, called 'terrace samples', we
use shadow evaporation and a resist bridge to obtain different
thicknesses for the leads and the constriction (see Fig.
\ref{sample}a). In the constriction, 15 nm Au is evaporated, while
for the leads 150 nm Au on top of 2 nm Cr is used. To decrease the
constriction length, an extra 50 nm Au is evaporated. This way, a
constriction of about $250 \times 100 $nm (length $\times$ width)
is obtained. Due to thick gold leads, the total resistance is only
30 $\Omega$ (including the wires of the measurement setup). For
the second kind of samples, called 'bow tie samples', we define a
constriction in the shape of a bow tie with a minimum width of 20
to 60 nm. In this case, the thickness of the gold in the leads and
the constriction is the same, namely 17 nm. Furthermore, a 3 nm
adhesion layer is used for most of the gold structure. By
evaporating this layer at an angle, we make sure that the Chromium
does not reach the constriction. Therefore, the very center of the
constriction contains gold only.
\\
Electromigration is performed by applying a slowly increasing
voltage (Keithley 230) to the wire, while monitoring the current
(Keithley 6517A). Measurements were done at room temperature (in
air) and at both 77 and 4.2 K (in cryogenic vacuum). After break
down, tunnel currents were determined. In our setup, zero-bias
tunnel resistances up to $10^{12}$ $\Omega$ can be measured.

\section{results and discussion}
\subsection{Excessive heating during electromigration}
\begin{figure}[h]
\begin{center}
\includegraphics[width=8cm]{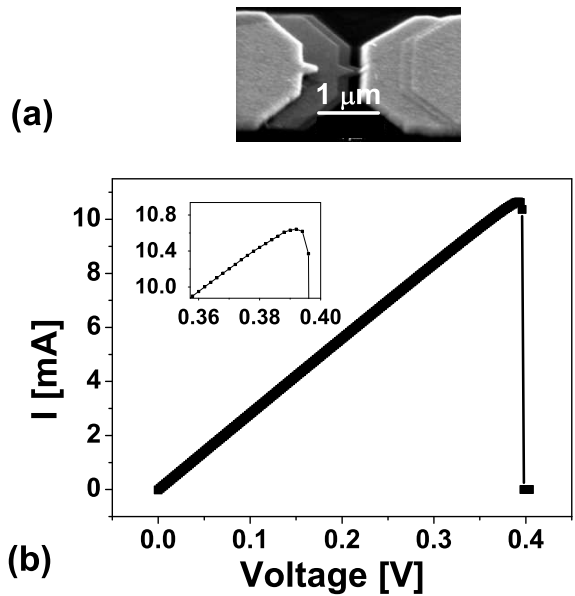}
\end{center}
\caption{a) Scanning electron micrograph of an electromigration
sample, 'terrace' type, made with electron beam lithography. The
thickness of the gold layer is 15 nm (middle), 50 nm, and 150 nm
(leads). The constriction in the middle is about 100 nm in width.
b) Representative breaking curve. The voltage is increased by 2
mV/s, while the current is observed. At 0.38 V, the
electromigration process starts, leading to a decrease in current.
At 0.4 V, after break down, a tunnel current can be measured of
about 500 nA. Inset: magnification of region just before break
down. Due to temperature increase and electromigration, dI/dV
decreases slowly, until final break down occurs. Measurement at 77
Kelvin, the initial total resistance is 36 $\Omega$. }
\label{sample}
\end{figure}

A typical graph of an electromigration experiment is shown in
Figure \ref{sample}b. Here, breaking is performed for a 'terrace
sample' (Fig. \ref{sample}a) by slowly increasing the voltage by 2
mV/s, at 77 K. The initial total resistance (including wiring) is
36 $\Omega$. At about 0.38 V, the current density has reached a
value that causes substantial mass flux, starting the formation of
a slit. At 0.4 V, the wire finally breaks down. Subsequently, a
small tunnel current can be observed, indicating that the gap size
is around 1 nm.

\begin{figure}[h]
\begin{center}
\includegraphics[width=8cm]{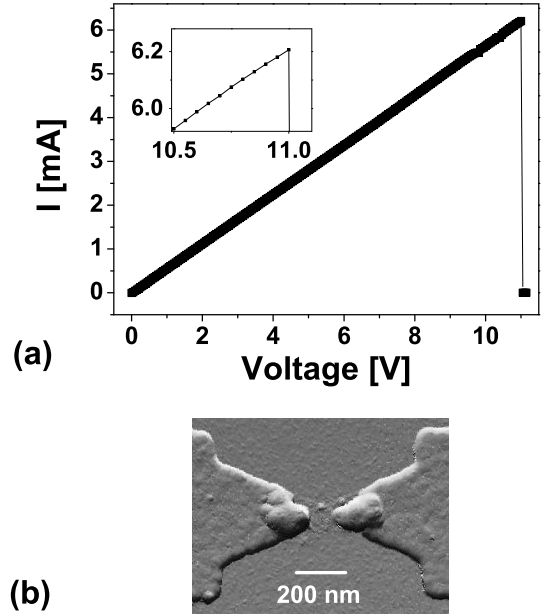}
\end{center}
\caption{a) I(V) measurement during an electromigration experiment
('bow tie' sample). The total resistance before electromigration
is 1780 $\Omega$. At 11 V, the sample breaks down abruptly. No
tunnel current could be observed afterwards. Inset: Magnification
of region just before breaking. Measurement at 295 K. b) AFM
picture of the sample after breaking.  As can be seen, the
breaking mechanism for this sample was melting instead of
electromigration. In between the electrodes, gold islands can be
observed. } \label{melting}
\end{figure}

A very different result is obtained for the experiment in Fig.
\ref{melting}. For this measurement, performed on a 'bow tie'
sample, we added an extra series resistance. The total initial
resistance equals 1780 $\Omega$. Although the graph in Fig.
\ref{melting}a) looks similar to the one in Fig.\ref{sample}b),
there are some obvious differences. First, the breaking process is
much more abrupt (see insets). Second, the voltage at break down
is much higher (11 V vs. 0.4 V). The most dramatic difference is,
however, that no tunnel current could be detected after break
down. To investigate the gap in more detail, we used atomic force
microscopy (AFM).  In Figure \ref{melting}b an AFM picture is
shown of the 'bow tie' sample after the breaking process. The
final gap size is approximately 100 nm, and the electrodes have
clearly rounded off. Most likely this is due to local melting.
Furthermore, there is a set of gold islands in between the
electrodes. We note that the presence of such nanoparticles may
lead to Coulomb blockade effects.\cite{sordan, hubert, gonzalez,
houck} From Figure \ref{melting}b we conclude that the break down
process has taken place in an uncontrolled manner, leading to high
local temperatures (and fields). We shall show below, that the
breaking dynamics is strongly correlated with the resistance in
series with the constriction, $R_s$. We note that this resistance
constitutes the main difference between the experiments in Figures
\ref{sample} and \ref{melting}. Minimizing $R_s$, results in a
much better control of the final gap size, making a feedback
system unnecessary.

\begin{figure}[h]
\begin{center}
\includegraphics[width=5cm]{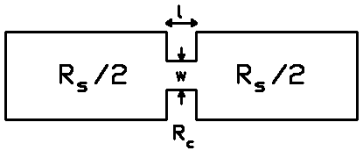}
\end{center}
\caption{ Schematic presentation of an electromigration sample.
The total resistance R consists of the constriction resistance
$R_c$ and series resistance $R_s$. Due to electromigration, the
constriction will shrink in time, leading to a reduction of the
width w and height h and an increase of $R_c$.}
\label{modelsample}
\end{figure}

To demonstrate the significance of $R_ s$, we use a simple model
for the constriction (a schematic drawing is shown in Fig.
\ref{modelsample}). We consider a slit of length l. Its width w(t)
and height h(t) are decreasing functions of time, due to gap
formation. The total resistance, $R_{tot}$, is the sum of the
constriction resistance, $R_c(t)=\frac{\rho  l}{h(t) w(t)}$, and a
series resistance $R_s$. In a virtual experiment, we ramp up the
voltage until electromigration begins at a bias $V=V_{c}$ (at $t
\equiv 0$). A more accurate way of saying that 'electromigration
begins' is to say that the flux of gold atoms, $J_m$, reaches a
certain critical value $J_m^c$.\cite{speed} For this to happen,
both the current density $j$ and the mobility (related to
$L^{*}_{m,m}$ in eq. (\ref{fluxmetal})) need to be substantial.
The mobility is strongly affected by the local temperature, which
increases due to Joule heating via $p=\rho
 j^{2}$. Consequently, electromigration does not start, until
a certain combination of current density $j_c$ and local
temperature $T_{start}$ is reached. The latter is directly related
to the local dissipation $p_c=\rho  j_c^{2}$ (below, we discuss
this in more detail).\cite{1D} The voltage $V_c$ and local
dissipation $p_c$ are related by $ p_c=\frac{V^{2}_{c}}{\rho
l^{2}}/(1+\frac{R_s}{R_c(0)})^{2}$. Next, we consider the breaking
process for $t>0$, keeping the voltage constant at $V=V_{c}$. Due
to electromigration a slit is formed in the constriction, leading
to an increasing constriction resistance $R_c(t)$. Consequently,
the local dissipation also increases, according to (in the
diffusive regime):

\begin{equation}\label{power}
    p(t)=p_c  \left (\frac{1+ \frac{R_s}{R_c(0)}}{1+\frac{R_s}{R_c(t)}} \right )^{2}
\end{equation}

Let us first consider the limit with a small series resistance,
$\frac{R_s}{R_c(0)} << 1$. In this case, the dissipated power is
always equal to $p_c$, independent of the width of the
constriction. Hence, the local temperature stays close to
$T_{start}$ and the gap grows slowly in time. In the other limit,
$\frac{R_s}{R_c(0)} >> 1$, the situation is different. At the
start of the process (t=0), the local power equals $p_c$. However,
the dissipated power increases rapidly during the breaking
process, especially when $R_c(t)>>R_c(0)$. In this limit,
$p>>p_c$. Consequently, uncontrollable heating takes place. This
can finally lead to melting of the electrodes.\cite{ivsource} We
note that equation (\ref{power}) only holds in the diffusive
regime, where the mean free path of the electrons is much smaller
than the dimensions of the contact. Interestingly, as the
electromigration process continues, the size of the constriction
decreases. Provided that heating is limited, one expects a
cross-over from the diffusive to the ballistic transport regime.
We will return to this phenomenon in section C.

To have a better control of the final gap size, one has to limit
the series resistance within the set-up as well as within the
sample itself. For the latter we developed the 'terrace' samples
that connect big leads to the constriction (see Fig.
\ref{sample}a). As compared to the 'bow tie' samples (see Fig.
\ref{melting}b), this geometry has three advantages. The most
important one is to limit the dissipated power during
electromigration, by the reduction of $R_s$. Another advantage is
that the voltage needed to break the samples is limited. This is
important, since, after break down all the voltage drop will be
across the gap. This leads to a high electric field. It is well
known that electric fields exceeding 2 V/nm, can cause a
reorganization of the electrodes. Depending on the shape of the
electrodes, this reorganization can result in a larger gap
size.\cite{mendez, huang} A third advantage of thick leads, is
their use as heat sinks. In this way, the heat can easily flow
away from the constriction. By the evaporation of 150 nm Au for
the leads, we have a total resistance of typically 30 $\Omega$
(including the wiring of the measurement setup).

\begin{figure}[h]
\begin{center}
\includegraphics[width=8cm]{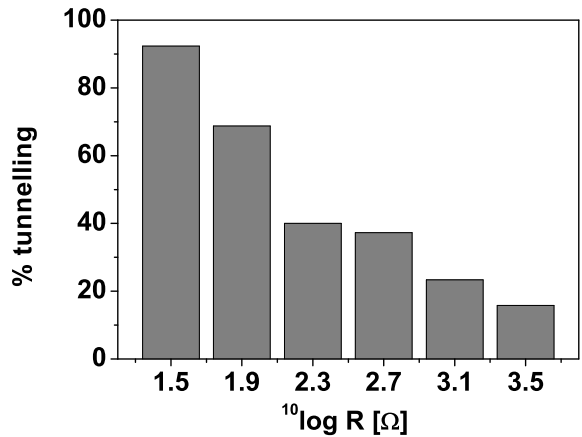}
\end{center}
\caption{Percentage of samples with a zero-bias tunnel resistance
$<$ $10^{12}$ Ohm after electromigration, as a function of the
initial series resistance. High initial series resistances lead to
excessive heating, causing large gaps. The number of samples
showing a tunnel current can be increased from 15 up to $>90$
percent by limiting the series resistance. In total, 142 samples
were measured. The number of samples per bar is 13/16/5/59/30/19.
} \label{histogram}
\end{figure}

To show the effect of a series resistance experimentally, a
variable resistance is inserted in series with the sample. This
way, we can check the final gap size as a function of $R_s$. The
result is shown in Fig. \ref{histogram}. We have investigated 142
samples, of both types. The wires were broken by slowly increasing
the voltage until a gap occurs. Afterwards, the tunnel resistance
at zero bias was measured, in order to make an estimation of the
gap size. Zero bias resistances larger than $10^{12} $ $ \Omega$
could not be measured, and are counted as "no tunnelling". As can
be seen from Fig. \ref{histogram}, there is a strong relation
between the gap size and the total initial resistance. Considering
samples with large series resistances, from $2-8$ $k\Omega$, only
15 percent have nanogaps with a measurable tunnel current. Samples
with low series resistances, however, give smaller gaps.
Decreasing the value for the series resistance to about $30$
$\Omega$, gives nanogaps with a measurable tunnel current in more
than 90 percent of the cases. From the tunnel current, an
estimation can be made of the size of the gaps. These are in the
range of 0 to 2 nm.

\subsection{Local temperature during electromigration}

For the model discussed above, we assumed that gap formation does
not start until a certain critical temperature $T_{start}$ is
reached. To test if this is indeed the case, we performed
measurements at 4.2, 77 and 295 Kelvin. We found no clear
differences in final gap size. We did observe, however, that a
higher critical power $p_c$ is needed, when samples are broken at
lower temperatures. This suggests that the local temperature at
which electromigration is triggered, is more or less independent
of the surrounding temperature. This is as expected, since the
mobility of gold atoms is strongly dependent on the temperature
(we will discuss this in more detail below).

In order to find the local temperature during electromigration,
T$_{em}$, we fabricated four terminal devices. Knowledge of
T$_{em}$ is important, since, for molecular electronics, molecules
are often put on the sample prior the breaking process. High
temperatures could damage the molecules, right where the nanogap
will be formed.

\begin{figure}[h]
\begin{center}
\includegraphics[width=7cm]{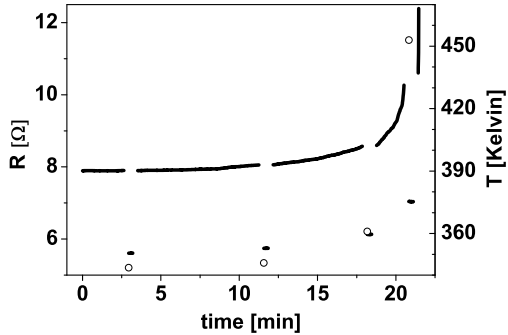}
\end{center}
\caption{Four-terminal resistance during electromigration. The
voltage is set for an initial electromigration speed of 100 $\mu
\Omega$/s, and kept constant afterwards. Due to electromigration,
grain coalescence and heating, the resistance increases.  Every
couple of minutes, the resistance is measured at low voltage, and
then V is set back to the initial value. From the change in
resistance between low and high voltage, the average local
temperature can be obtained (circles). The temperature of the
environment is 77 K, while the temperature at which
electromigration is initiated is 345 K. Just before break down,
the temperature increases. The last temperature measured before
break down is 450 K. The applied voltage is 0.72 V. The initial
two-terminal resistance is 41 $\Omega$, the four-terminal
resistance is 5.8 $\Omega$ (at 77 K). The initial current density
is 4$\times 10^{8}A/cm^{2}$.} \label{temp}
\end{figure}

To determine the local temperature, the samples were first cooled
down to obtain the relation between temperature and (local)
resistance (4.2 K $<$ T $<$ 295 K).\cite{lambert, dumpich,durkan}
The electromigration experiment is subsequently performed as
follows. First, the voltage is increased till electromigration
starts (defined as $dR/dt = 100 \mu \Omega$/s), and kept constant
afterwards. An example is shown in Fig. \ref{temp}. From here on,
different processes will influence the local resistance. The most
important ones are electromigration, heating and the coalescence
of grains. To make an estimation of T$_{em}$, it is crucial to
deduce only the resistance change due to heating. This was done by
interrupting the measurement every couple of minutes, and
determining the resistance at low voltage.\cite{dumpich} Finally,
the average local temperature T$_{em}$ is estimated from the
resistance difference at low and high voltage and the
(extrapolated) relation between temperature and resistance.

For the trace in Fig. \ref{temp}, the initial four-terminal
resistance was 5.8 $\Omega$ (at 77 Kelvin). Setting the voltage to
start the breaking process, increases the resistance drastically
to about 8 $\Omega$ (due to heating). Subsequently, the resistance
increases in time, and accelerates just before breaking.
Furthermore, the local temperature increases, as predicted by eq.
(\ref{power}). We note that there is always some series
resistance, so that heating cannot be taken away completely.
However, when high series resistances are used, the temperature
does not only reach higher values, it also diverges much faster.
This is clearly shown in the insets of Figures \ref{sample}b and
\ref{melting}a. In Fig. \ref{sample}b, a slow increase in
differential resistance is observed. In contrast, for Fig.
\ref{melting}b, gap formation goes all at once, and the breaking
process cannot be resolved.

\begin{table}
\caption{\label{tabel} Local temperature at which electromigration
starts, $T_{start}$, for 5 different samples. Interestingly,
$T_{start}$ appears rather independent of the temperature of the
sample surroundings, $T_0$.  For data set C and E, the temperature
is also measured just before break-down, $T_{final}$. This
temperature depends strongly on to what extent the process is
controlled. Data set C is taken from Fig. [\ref{temp}].}
\begin{ruledtabular}
  \begin{tabular}{cccc}
   ~~Data set~ & $~~T_{0} [K]$ ~& ~$T_{start} [K]$ ~&~$  T_{final} [K] $~ \\
  \hline
  A & $4.2$ & $430 \pm 40$& $-$  \\
  B & $4.2$ & $440 \pm 40$& $-$  \\
  C & $77$  & $345 \pm 10$& $455 \pm 10$  \\
  D & $295$ & $380 \pm 40$& $-$  \\
  E & $295$ & $420 \pm 10$& $460 \pm 10$  \\

 \end{tabular}
 \end{ruledtabular}
\end{table}

In Fig. \ref{temp}, the local temperature increases from 77 K to
345 K, when the voltage is set to start the breaking process. The
temperature determined just before break down, $T_{final}$, is
about 450 Kelvin. In Table [\ref{tabel}], four other data sets are
shown, at different environment temperatures. The temperatures at
which electromigration is triggered are close together, whereas
the variation in the temperature of the environment is about 290
Kelvin. This is a strong indication that gap formation is indeed
due to electromigration (at a critical temperature) and not due to
thermodiffusion via a temperature gradient.\cite{durkan} \\
To understand why $T_{start}$ is rather independent of the
temperature of the sample environment, $T_0$, we introduce a
simple model. To fit its parameters, we use the data taken at 77 K
(Fig. \ref{temp}). We rewrite eq. (\ref{fluxmetal}), by noting
that $\nabla \mu^m = \Omega d \sigma / dx$ is the driving force
due to stress $\sigma$:\cite{thin film,blech}

\begin{equation}\label{Au1}
    \textbf{J}_m=-L^{*}_{m,m}(\Omega \Delta \sigma / L - Z^{*}e\rho j)
\end{equation}

Here $ \Delta \sigma$ is the stress built up over the length L. As
long as $\Delta \sigma<\Delta \sigma_{max}$, the electromigration
force is balanced by the stress gradient and the atom flux $J_m$
is zero.\cite{thin film,blech} This defines a critical current
density $j_{min}=\frac{\Omega \Delta \sigma_{max}}{Z^{*}e \rho
L}$, which is independent of temperature, since $Z^{*} \rho$ is
temperature independent.\cite{sorbello, verbruggen} Using the
Einstein relation $D=L^{*}_{m,m} d \mu / dc $, where D and c are
the diffusion constant and concentration respectively, we can
write down the temperature dependence of the atom flux. For the
chemical potential we have $\mu = k T$ln$ c$,\cite{thin film} and
for the diffusion constant $D= D_0 e^{-E_a/kT}$ so that we can
write for the phenomenological coefficient:
$L^{*}_{m,m}=\frac{D_0}{k T} c e^{-E_a/kT}$. Here, $E_a$ is the
activation energy of gold diffusion on the surface, which is 0.12
eV.\cite{Wilson} This leads to the following approximation for the
atom flux due to the current density:

\begin{equation}\label{Au3}
    \textbf{J}_m=\frac{\alpha}{T} (j-j_{min})e^{-E_a/kT}
\end{equation}

where $\alpha \equiv c D_0 Z^{*} e \rho /k <0$. For our
measurements we use temperatures in between liquid Helium and Room
temperature and the applied voltage over the constriction is
typically 20 mV or higher. In this range, the inelastic scattering
length of the electrons does not exceed the contact size.
Therefore, the effective temperature in the contact can be
described by $T= T_{0}+\beta j^{2}$, where $\beta$ is a fit
parameter.\cite{beta} We determine $\beta$ from the first
temperature point in Fig. \ref{temp}: $T_{start}=345$ K at $j=4
\times 10^8 A/cm^2$. Furthermore, we estimate that $j_{min}$
equals $j_{min} \approx 1\times 10^{8} A/cm^{2}$.\cite{thin film}

\begin{figure}[h]
\begin{center}
\includegraphics[width=7cm]{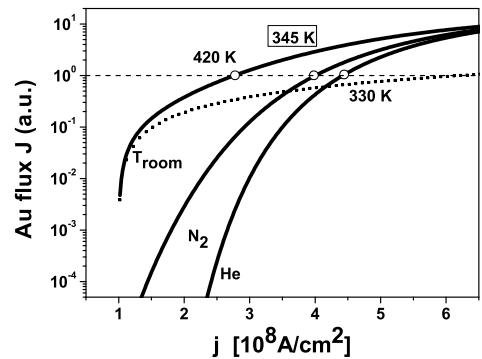}
\end{center}
\caption{Mass flux $J_m$ (arbitrary units) versus current density
j, for various surrounding temperatures $T_0$ (see eq.
(\ref{Au3})). From Fig. [\ref{temp}], we know that
electromigration starts (i.e., that $J_m=1$) when $j=4 \times 10^8
A/cm^2$ with a local temperature $T_{start}$= 345 K (rectangle).
We use this to obtain the parameters $\alpha$ and $\beta$ in eq.
(\ref{Au3}). With these, $J_m$ versus j, can also be plotted for
$T_0=4.2$ K and $T_0=295$ K, respectively. Taking Joule heating
into account (solid lines), the temperature at which
electromigration starts is 330 K for $T_0=4.2$ K and 420 K for
$T_0=295$ K, respectively. Note that when Joule heating is
neglected, electromigration can only set in at room temperature
(dotted line). For 4.2 and 77 K, the critical mass flux ($J_m=1$)
cannot be reached for reasonable j.} \label{model}
\end{figure}

Equation [$\ref{Au3}$] implies a strong relation between
temperature and mass flux, especially due to the activated
behavior. Gap formation starts as soon as the mass flux is large
enough, i.e., when $J_m$ reaches a certain value $J_m^c$. We
define this quantity, in arbitrary units, as $J_m^c \equiv 1$.
Going back to Fig. \ref{temp} with $T_0=77$ K, we infer that for
$j=4 \times 10^8 A/cm^2$, we have $J_m=J_m^c=1$. From this, we
deduce the other parameter $\alpha$. Knowing both $\alpha$ and
$\beta$, we can plot $J_m$ as a function of current density j for
all three experimental temperatures $T_0$. This is shown in Fig.
\ref{model}. For each $T_0$, we calculate the current density and
temperature at which electromigration sets in, $T_{start}$,
demanding $J_m=1$. (For $T_0=77$ K, this is by definition at $j=4
\times 10^8 A/cm^2$ with $T=345$ K). For $T_0=4.2 K$ and $T_0=295
K$ we find that $T_{start}$ equals 330 and 420 K, respectively.
Hence, increasing the temperature of the environment by 290 K,
increases $T_{start}$ by only 90 K. Related to this is the fact
that the critical current density increases with decreasing $T_0$.
Although basic, the model is in rather good agreement with the
values of Table [\ref{tabel}].

Do previously applied molecules get damaged at these temperatures?
As we have shown above, the local temperature just before break
down depends on the exact sample design. Also, the maximum
temperature molecules can stand, strongly depends on the molecule.
At 460 Kelvin, which is $T_{final}$ in Table [\ref{tabel}], most
molecules used in molecular electronics, do not get damaged.
However, care should clearly be taken. The electromigration
process should be controlled either by optimizing the sample
design, or by using active feedback. If not, the local temperature
increases dramatically as the wire gets thinner and may reach
values up to the melting point of gold.

\subsection{Quantized conductance}

\begin{figure}[h]
\begin{center}
\includegraphics[width=7cm]{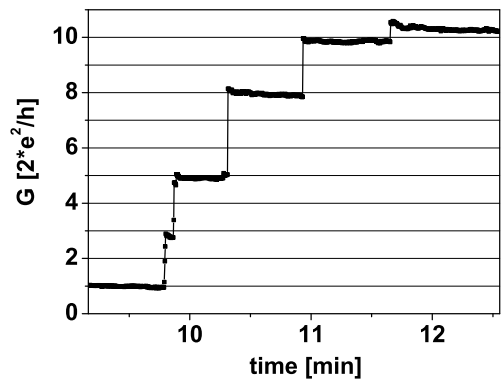}
\end{center}
\caption{Quantized conductance during the electromigration
procedure at constant voltage. The voltage is set for an initial
electromigration speed of 100 $\mu \Omega$/s, starting the
breaking process. The resistance increases due to electromigration
until, after 9 minutes, a conductance of 1 $\times$ $2e^2/h$ is
observed. After another minute, the configuration starts to
reorganize, leading to a stepwise increase in conductance.
Measurement at room temperature with $V=0.73$ V and
P$=1\times10^{-6}$ mbar.}\label{qc}
\end{figure}

As we have shown, decreasing the series resistance gives a better
control of the final gap size. Another interesting consequence of
controlled electromigration, is the possibility to observe the
transition between the diffusive and the ballistic
regime.\cite{strachan} If slit formation is a well-defined
process, at some point the constriction gets smaller than the
inelastic scattering length. In this case, conduction becomes
ballistic. This is indeed what we observe in a number of cases.
During the breaking process, the sample conductance does not go
smoothly to the tunneling regime, but locks in at plateaus equal
to integer values of the conductance quantum $2e^{2}/h$
.\cite{bart,muller} An example is shown in Fig. \ref{qc}. Here, we
start with a resistance of about 45 $\Omega$, increase the voltage
until we see a change in conductance and, from there on, keep the
voltage constant. Slowly, the conductance goes down, to finally
stop at 1$\times 2e^{2}/h$. At this point, the electromigration
process stops automatically, due to the almost perfect
transmission of the gold atoms. The corresponding current density
is about 57 $\mu A / $atom. This is close to the maximum value
described elsewhere,\cite{mizobata} and is much larger than the
maximum current density in the diffusive regime. Nevertheless, the
electrodes reorganize due to atomic diffusion. Slowly, the contact
evolves from a single atom contact to a contact with 3, 5, 8 and
10 gold atoms.\cite{statistiek} At much higher voltages, it is
still possible to break the constriction and obtain a tunnel
resistance. We note that we have only observed quantized
conductance in samples with minimized $R_s$. If the power is
unlimited, the high temperature prohibits the formation of a
single atom contact.

\section{conclusions}

To conclude, we have investigated the role of dissipation during
the formation of nanogaps due to electromigration. We find the
following. First, some Joule heating is needed for gap formation
to begin. The process does not start until a local temperature of
typically 400 K is reached. This value is rather independent of
the temperature of the surroundings. We relate this phenomenon to
the activated behavior of the atomic mobility, which plays a
crucial role in this diffusion process. Second, although Joule
heating is important to start electromigration, it can also lead
to unwanted effects. If no measures are taken, the temperature in
the constriction can increase up to the melting point of gold,
during slit formation. This leads to large gaps, possibly
containing small gold islands. By minimizing the total series
resistance of the system, we make sure the temperature keeps low
during the electromigration process. In this way, we increased the
number of samples with a gap $<$ 2 nm to $>$90 percent, without
the use of active feedback. With the improved samples, gap
formation is much slower. Hence, it is also possible to observe
the transition from the diffusive to the ballistic regime. In a
number of cases, contacts have been achieved that consist of only
a single or a few atoms.

\begin{acknowledgments}
We thank Gert ten Brink, Niko Tombros, Bernard Wolfs and Siemon
Bakker for their assistance and advice. We thank Herre van der
Zant and Hubert Heersche for useful discussions. This work was
financed by the Nederlandse Organisatie voor Wetenschappelijk
Onderzoek, NWO, via a Pionier grant.
\end{acknowledgments}

\newpage

\renewcommand{\baselinestretch}{2}


\begin{thebibliography}{9}
\bibitem{speed}
For this manuscript we define the starting point of gap formation,
when the increase in resistance reaches a speed of 100 $\mu
\Omega$/s. We note that for other values, the starting temperature
$T_{start}$ will also change.
\bibitem{1D}
Since we use a 1-dimensional model, the length of our constriction
does not change with time.
\bibitem{ivsource}
A comparison can also be made with a current source (Norton
equivalent circuit) and a voltage source (Thevenin circuit). Gold
wires with high series resistances are broken as if connected to
an ideal current source. As the width of the wire decreases, the
current density increases. Constrictions with negligible series
resistances are broken as if connected to an ideal voltage source.
As the width decreases, the current density stays approximately
constant.
\bibitem{beta}
In our measurement range, the applied bias voltage is close to the
Debye energy, leading to strong electron-phonon scattering. The
inelastic scattering length does not exceed the contact size and
the temperature in the contact can be described by the given
relation. However, When temperatures around or below 1 Kelvin are
considered, and the maximum voltage is in the mV range,
electron-electron scattering is the dominant relaxation mechanism.
The relation for the temperature becomes $T= \sqrt{T_{0}^2+\beta
j^{2}}$. For the measurement at Helium temperature, however, this
will give only a minor change in the final maximum temperature
(343 K instead of 330 K).
\bibitem{statistiek}
Since it is not possible to vary the conductance back and forth,
as with the mechanically controllable break junctions, we have not
performed statistics. However, since the values are exactly equal
to the conductance quantum,  there is a strong indication that the
conductance steps are due to single gold atoms.
\end{thebibliography}
\end{document}